%% file: benchmark.tex
\documentclass{sig-alternate}

\usepackage{comment}
\usepackage{amsmath}
\usepackage{amssymb}
\usepackage{cite}
\newtheorem{example}{Example}
\usepackage{enumerate}
\usepackage{enumitem}
\setlist{leftmargin=1.5em, itemsep=1pt, topsep=3pt}

\newcommand{\onefig}[3]{
 \begin{figure}[!h]
 \vspace{-0.1in}
  \centering % SJ: Apparently, \centering is more space efficient that \begin{center}
    \includegraphics[width=#3]{#1}
    \vspace{-0.1in}
    \caption{\label{#1}#2}
    \vspace{-0.15in}
 \end{figure}
}

\newcommand{\onefigtop}[3]{
 \begin{figure}[!t]
 \vspace{-0.1in}
  \centering % SJ: Apparently, \centering is more space efficient that \begin{center}
    \includegraphics[width=#3]{#1}
    \vspace{-0.1in}
    \caption{\label{#1}#2}
    \vspace{-0.15in}
 \end{figure}
}

\newcommand{\onefigfullrow}[3]{
 \begin{figure*}[!ht]
 \vspace{-0.1in}
  \centering 
    \includegraphics[width=#3]{#1}
    \vspace{-0.24in}
    \caption{\label{#1}#2}
    \vspace{-0.15in}
 \end{figure*}
}

\newcommand{\hakanname}{{\fontencoding{T1}Hakan Hac{\i}g\"{u}m\"{u}\c{s}}}

\renewcommand{\baselinestretch}{0.980}

\begin{document}
%
%% --- Author Metadata here ---
%\conferenceinfo{WOODSTOCK}{'97 El Paso, Texas USA}
%%\CopyrightYear{2007} % Allows default copyright year (20XX) to be over-ridden - IF NEED BE.
%%\crdata{0-12345-67-8/90/01}  % Allows default copyright data (0-89791-88-6/97/05) to be over-ridden - IF NEED BE.
%% --- End of Author Metadata ---

\conferenceinfo{DanaC'13,} {June 23, 2013, New York, NY, USA} 
\CopyrightYear{2013} 
\crdata{978-1-4503-2202-7/13/6} 
\clubpenalty=10000 
\widowpenalty = 10000

\title{Towards a Workload for Evolutionary Analytics} 

%%\subtitle{[Extended Abstract]
%\titlenote{A full version of this paper is available as
%\textit{Author's Guide to Preparing ACM SIG Proceedings Using
%\LaTeX$2_\epsilon$\ and BibTeX} at


%\numberofauthors{0} %  in this sample file, there are a *total*
\author{
 \alignauthor Jeff LeFevre\raisebox{1.5mm}{$^{+*}$}~~~~~Jagan Sankaranarayanan\raisebox{1mm}{$^{*}$}~~~~~\hakanname\raisebox{1.5mm}{$^{*}$} \\ Junichi Tatemura\raisebox{1.5mm}{$^{*}$}~~~~Neoklis Polyzotis\raisebox{1.5mm}{$^{+}$}\\
 \raisebox{1.5mm}{$^*$}NEC Labs America, Cupertino, CA~~~\raisebox{1.5mm}{$^+$}University of California Santa Cruz\\
 \vspace{0.1in}
 \fontsize{9}{9}\selectfont\ttfamily\upshape{\{jlefevre,alkis\}@cs.ucsc.edu, \{jagan,hakan,tatemura\}@nec-labs.com}\\
}

\maketitle
\input{abstract}

%A category including the fourth, optional field follows...
%%\category{D.2.8}{Software Engineering}{Metrics}[complexity measures, performance measures]
\vspace{-.1in}
\vspace{-.1in}
%\keywords{Databases, workloads, analytics, benchmarks, metrics}
%\keywords{Databases, big data, workloads, analytics, benchmark, metrics}

\input{introduction}

\input{sec-two}

\input{sec-three}

\input{sec-four}

\input{sec-five}

\vspace{-0.12in}
\bibliographystyle{abbrv}
{
\renewcommand{\baselinestretch}{0.95}
\small
\vspace{0.05in}
\bibliography{benchmark}  % sigproc.bib is the name of the Bibliography in this case
}
% ACM needs 'a single self-contained file'!
\balancecolumns
%\appendix 
\input{appendix}

%\balancecolumns % GM June 2007
% That's all folks!
\end{document}

%% file: abstract.tex
\begin{abstract}
Emerging data analysis involves the ingestion and exploration of new data sets,
application of complex functions, and frequent query revisions
 based on observing prior query answers. 
We call this new type of analysis evolutionary analytics and identify its properties.
This type of analysis is not well represented by current benchmark workloads.
In this paper, we present a workload and identify several metrics 
to test system support 
for evolutionary analytics.
Along with our metrics, we present methodologies for running the 
workload that capture this analytical scenario.

\end{abstract}

%% file: introduction.tex
\section{Introduction}\label{sec-introduction}

A new analytical landscape has emerged, exemplified by the popularity of 
``big data''  systems such as Hadoop as well as the recently
added support for big data processing by all major 
data warehouse vendors~\cite{groe2011, hens2011c}.
%~\cite{groe2011,hens2011a,hens2011b,hens2011c}.
Data volumes are growing rapidly and log files are an important data
source, e.g., social media or sensor data.
Queries are often exploratory in nature, and  
system-facilitated data exploration has been proposed 
in~\cite{sell2013,thus2010,khou2009,chat2009}.
Given this scenario, new requirements for analysis have been noted in~\cite{cohe2009}, 
including the need to access ``disparate, decentralized data''~\cite{fran2005}.
% and apply complex processing methods.
Analysis frequently includes complex processing methods such as user defined functions (UDFs), 
e.g.,~\cite{hell2012,mahout, datafu},
 created by expert users for domain-specific processing needs.

%others~\cite{hadepot,weik2013} have expressed the desire for realistic,
%non-trivial workloads and datasets in this space.
%This landscape is exemplified by ``big data'' and big data 
%processing systems
%such as Hadoop, as well as the recent extension of many commercial data 
%warehouses to include support for  
%The volume of data is growing rapidly, and new data sources include 
%log files from social media or sensor data. 
%Log schemas may evolve over time and log data is frequently incomplete or noisy.
%, requiring some amount of transformation prior to analysis.
%often this log data is incomplete or noisy.

%Current analytic tasks frequently include , often created by expert users.
In this new analytical setting, data analysts and data scientists are becoming 
increasingly important to businesses~\cite{pati2011,bowi2012}.
%Expert users such as data analysts often create and apply user defined functions (UDFs) for domain-specific processing tasks.
%UDFs are common in the big data space, and recent work provides in recent work~\cite{hell2012,mahout, datafu}.
Because analysts are tasked with finding value within their growing data 
sources, the speed at which an analyst can iterate through successive 
investigations to gain insight is crucial~\cite{lava2011}.
%Queries can be exploratory in nature, and a query may change as the user
%gains better understanding of the data.
To measure system performance, there is a need for a workload and metrics 
to capture this emerging type of analytics.
%Current analytical benchmarks such as TPC-DS~\cite{tpc} do not adequately
%represent this type of analytical workload.
%This type of analysis requires flexible analytics systems, 
%Current analytical benchmarks such as TPC-DS~\cite{tpc} do not adequately
%represent this type of analytical workload.
%For instance, TPC-DS queries focus on known data and known reporting tasks, 
%with a carefully designed schema and data warehouse.
%This is not always possible in the current analytical scenario, as it may not
%afford the up-front, top-down design of a traditional data warehouse.
It is important to understand the features of this new type of workload and 
effective ways to evaluate system performance in this space.
%We term this scenario evolutionary analytics and identify the following three 
%characteristic properties.
We term this scenario \emph{evolutionary analytics}
and identify the following three important characteristics of evolutionary analytics
that are not captured by existing benchmarks.

  \begin{enumerate}
    \item \textit{Query Evolution}. 
    Queries are exploratory and evolve over time.
    A query may go through multiple evolutions (versions) whereby an 
    analyst iteratively formulates, tests, and refines hypotheses
    during investigation. 
    %We define a query version as a revision to the previous query, 
    %representing an analyst examining the output and reformulating the query.
    Query revisions appear as a sequence of mutations to the original
    query, and this temporal nature is a key feature.
    While traditional interactive OLAP may perform operations such as 
    roll-up or drill-down to slightly modify the query, revisions in 
    exploratory analysis can include more types of changes to the 
    query and a longer sequence of changes, 
    as we define in Section~\ref{sec-query-building-blocks}.
    %These represent the exploratory process.
    Typical revisions are minor refinements as well as more significant
     changes such as augmented functions, addition (or removal) of sub-queries,
     or incorporating new data sources to obtain richer answers.
    %As a query evolves, a new version of a query may have significant overlap 
    %with a previous version of a query.
    %We term this feature \textit{Query Evolution}.
    
    \item \textit{Data Evolution}.
    %ETL on-the-fly aspect.  
    Queries may incorporate new data from external sources such as raw logs 
    or local data files.    
    New data sources should be easily ingested or accessible for use 
    during query processing, and these data sources may have evolving schemas.
    A formal ETL (extract, transform, load) project for a data warehouse can 
    have a very high cost in dollars and design time. %, as noted in~\cite{simi2009}.
    Enabling access to diverse data sources via ETL on-the-fly 
    is a key feature of this new analytical environment.

    \item \textit{User Evolution}.
    A flexible and accessible system should enable new users to get started 
    posing queries testing different hypotheses, potentially over old or new data sets.
    %Data sets are large and processing can represent a significant investment of system effort. 
    New users in the system arrive less frequently than query revisions,
    and their queries do not closely resemble another user's queries.   
     %  and the ability to share prior work across  users is a key feature.

    %re-use results and exploit prior computation.    
    %New users may pose queries testing different hypotheses over the 
    %same datasets, thus common processing tasks may be repeated. 
    %A new query may have some degree of overlap with other analyst queries.

  \end{enumerate}
  
  %\item Taken together, we term this Online Evolutionary Processing or 
  %\textit{OLEP} and these features differentiate it from OLTP and OLAP 
  %workloads in significant ways.
  %Moreover, queries may frequently include user-defined functions (UDFs)
  %for domain-specific processing tasks.
  %Thus UDFs are an important component of this analytical landscape.

In this paper, we propose a workload with these features and 
metrics to test how well a system supports them.
Query response time is a primary metric but it is useful to understand
system performance for other metrics as well.
For example, response time may hide several other system overheads, such as 
the overhead to tune the physical design (if this happens online as queries 
get executed) or the cost to load data (if it has to be ETL'ed on-the-fly).
By separating out these overheads in different metrics, we can see where each
system excels and also understand how to develop hybrid systems that
combine their best features.

\onefigtop{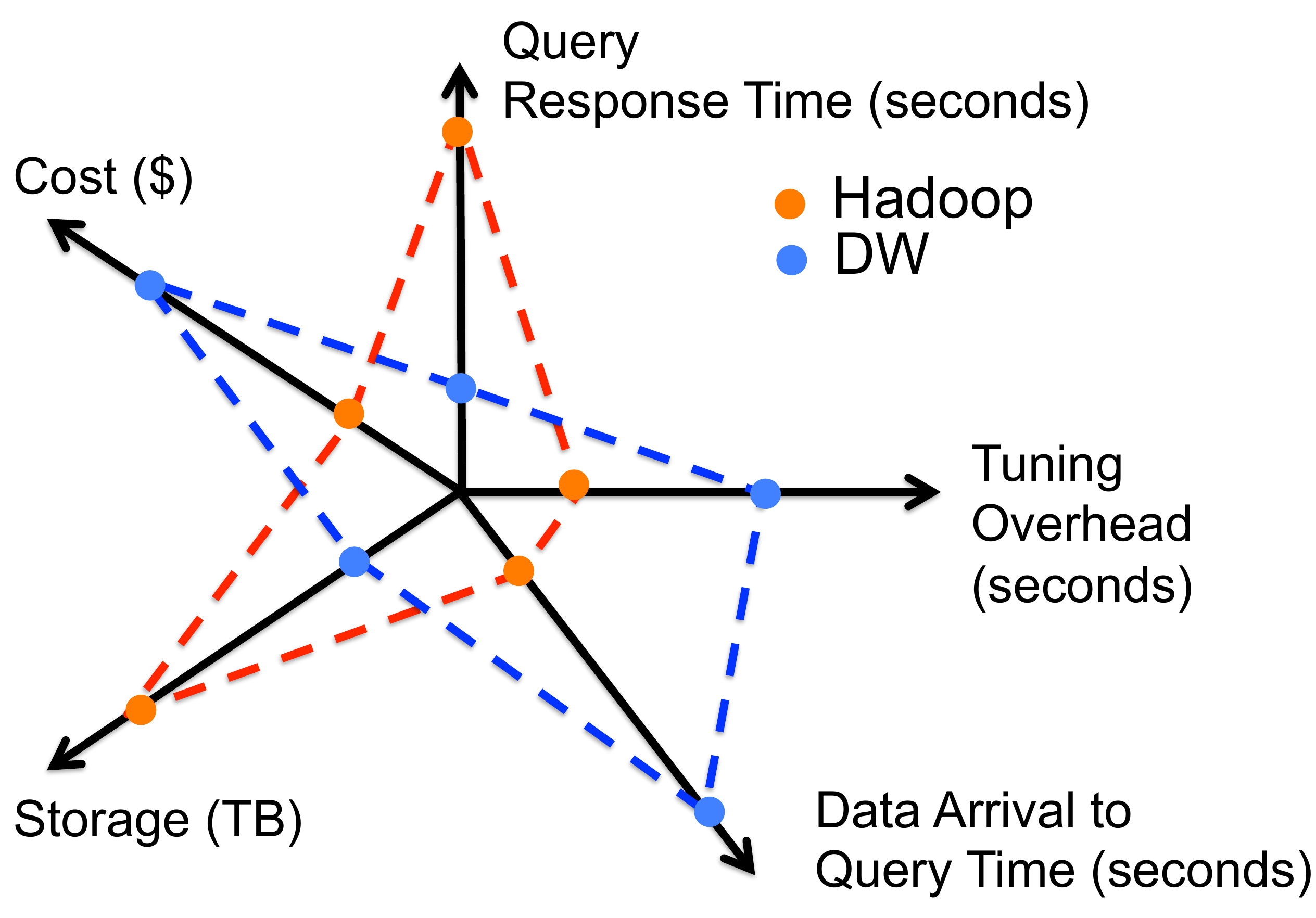}{System metrics for evolutionary analytics}{2.2in}

Figure~\ref{fig-evaluation-dimensions} shows our proposed metrics as dimensions
 (although not completely orthogonal).
Query response time indicates how quickly analysts can arrive at answers when 
testing hypotheses.
Tuning overhead represents the time expended for physical design tuning, 
i.e., creating indexes and materialized views, to improve query 
processing speed.
Data arrival to query time indicates the time until newly arrived data is 
query-able. 
Storage in terabytes indicates the overhead for all data and auxiliary data 
structures (indexes and views).
Cost in dollars represents the system cost to process the workload.
Along each dimension, we indicate the relative performance of a traditional data 
warehouse (\textsc{dw}) and a \textsc{Hadoop} system. % to illustrate their relative differences.
This illustration shows how to compare and contrast different systems using our proposed metrics.
%This illustration highlights the usefulness of each metric and contrasts
%different systems that choose various tradeoffs among these metrics to achieve 
%good performance.
We present experimental results for four data systems 
using these metrics in Section~\ref{sec-four}.

In this paper we make the following contributions.
  
  \begin{itemize}
    \item We define evolutionary analytics along with our notions of 
    query evolution, data evolution, and user evolution, and introduce 
    a workload with these properties.
    \item We propose relevant metrics for evolutionary analytics and describe 
    their tradeoffs.
    \item We show how metrics can be used to guide the design of hybrid systems that target the best 
    features of specialized processing engines.
    %why understanding the tradeoffs is important in this scenario.    
    % \item We identify properties of evolving queries, common changes observed empirically, and propose a workload with these properties.
    % \item We describe methodology for using this workload to measure system performance regarding evolving analytics.
  \end{itemize}

%% file: sec-two.tex
\section{Workload characteristics and system metrics}\label{sec-two}

In this section we first describe our workload properties and contrast
with other benchmarks, then we describe our metrics to test system
support for evolutionary analytics and highlight the various tradeoffs for 
each metric. 

\subsection{Workload Characteristics}

Data analysis queries dealing with low-structured or log data
must often perform data extraction tasks as well as analytical tasks, 
including the application of machine learning algorithms.
Some recent examples of such queries are given in~\cite{park2011,simi2012}.  
These queries reference Twitter data and static data such as IMDB or
historical business sales data.
They use several UDFs which perform sentiment analysis and classification 
tasks.
One query infers movie rating trends for two consecutive months,
and the other computes the impact of a  marketing campaign 
in different sales regions.
These queries are representative of current data processing tasks.

Given our previously described workload needs and examples from recent data analysis tasks, 
we determine the following desirable characteristics for the query workload
of the benchmark.
\begin{itemize}
  \item Significant query complexity, including common UDFs, 
  performing non-trivial analysis tasks to gain insights and find value 
  within unproven data sources.
  
   \item Several successive versions for each query, representing data 
  exploration during hypothesis testing and query refinement. 
  %%Revisions should include specific types of mutations only.

  \item Access realistic data sets during query processing.  
  These should include raw logs and static/historical data sets.
  %some of which may be flat files that are not pre-cleaned and loaded.
  %\item Incorporate UDFs common to this type of data analytics,
   %e.g., classifiers.
\end{itemize}

\begin{comment}
An important aspect of evolutionary analytics is the way in which the workload
is observed by the system.
The notion of query evolution requires successive versions of a query to appear
 in temporal sequence.
%This temporal workload feature is important to system performance since a new 
%query version may repeat many of the tasks from previous versions.  
The notion of data evolution requires the arrival of new data 
sets  during query processing that require ETL on-the-fly.
The notion of user evolution requires the introduction of new users 
throughout workload execution.
New users should arrive less frequently 
than reformulated queries, and their queries should not closely resemble 
another user's queries.
%In this respect the analytical tasks should be different, although they may
%incorporate some of the same data sets and similar processing tasks.
%The exploratory nature of evolving analytics makes 
%planning ahead difficult, so a system's ability to exploit prior computation 
%toward new purposes becomes important to support new users efficiently.
\end{comment}

Current analytical benchmarks such as TPC-H and TPC-DS~\cite{tpc} 
do not adequately capture this type of analytical workload.
For instance, TPC-DS queries focus on known data and known reporting tasks, 
with a carefully designed, fixed schema for a data warehouse.
This is not always possible in the current analytical scenario, as the use-case may not
afford the up-front, top-down design of a traditional data warehouse.
In contrast to query evolution where a query goes through an ordered
sequence of mutations, the set of reporting queries in TPC-DS represent 
independent tasks where the ordering of one query is not dependent on 
the previously executed query.
In contrast to data evolution,
TPC-DS maintenance workloads reflect table inserts from its 
counterpart OLTP database, but they do not reflect a growing log or 
arrival of a new data source.

\subsection{System Metrics}

We propose the following metrics to evaluate a system for evolutionary analytics. 

\paragraph*{Query response time} 
Query performance is a key metric of the benchmark, and measures total workload execution time.  
This metric serves as the primary indicator of how well a system is able to support 
the workload features and process the workload efficiently.
%One of the main objectives this metric is intended to capture in our scenario 
%is the ability of a system to reuse prior results effectively. 

\paragraph*{Tuning overhead}
Physical design tuning can greatly improve query performance.  
Tuning might be considered offline during a system maintenance window or online during workload processing.
This metric reports the cumulative time spent on tuning, which is
the time spent to run a tuning tool and the time to materialize all indexes and views.

\paragraph*{Data arrival to query time}
This metric reports the time until newly arrived data is available to query.  
%In our scenario this effort is no longer completely offline, as in traditional database design.  
Data preparation is an atomic operation that enables the data to be accessed by a query.  
This may include the schema definition such as a 
\textsc{create table} statement and a \textsc{load} operation.

\paragraph*{Storage size}
This metric indicates the total storage required in terabytes.
Total storage includes that required for all base data, and all indexes and materialized views.
Storage size can be asymmetrical even for base data, considering some 
systems replicate data by design (e.g., Hadoop). % whereas other systems might not.

\paragraph*{Monetary cost}
This metric indicates the total system cost for query processing and
data storage.
For simplicity, in this work we use dollar cost to include only machine time and storage cost.
A better cost metric could be total cost of ownership (TCO), which 
includes system administration cost as well as hardware cost.
The cost metric can guide tradeoffs that are tolerable for a given environment, 
i.e., exploratory analysis  where return on investment may not be known.

\subsubsection{Metric tradeoffs}

Our metrics can be used to understand the various tradeoffs to consider for system design.
Previous studies~\cite{pavl2009}  have considered load times and query response time.
Here we introduce additional metrics and show how they interact with each other.
%but here we include additional metrics we show the interaction 
%There are tradeoffs among each metric, and understanding these tradeoffs is beneficial
Clearly response time interacts with all of the other metrics of data loading, physical design tuning, storage space, and cost.
Reducing response time can be achieved through a combination of tradeoffs among the other metrics. 
 
For example, tuning overhead impacts both query response time and storage size.  
A good physical design can consume multiple times the size of the base data,
but may reduce workload cost dramatically.
Due to their size, the choice of indexes and views will also appear as a tradeoff along the storage metric.
Loading may require data cleaning, transformation, 
and copying/storing the data, which is a typical ETL task in a data warehouse.
In contrast, using Hive~\cite{thus2010} requires  only  the schema definition to be provided
before a query can access the data.
% which resides in its native files in the distributed file system.
This presents a tradeoff between query response time and data load time.
%if the high cost of loading is incurred vs. accessing the data in its native form.

The cost metric leads to interesting tradeoffs for system design.
In particular, the advent of the cloud enables pay-as-you-go performance,
allowing for a rich set of choices for query processing.
For example, Hadoop~\cite{amazon-aws, google-engine},
databases~\cite{amazon-aws, ms-azure}, 
and recently even petabyte-scale data warehouses (e.g.,Redshift~\cite{amazon-aws}) are all available on-demand.
Moreover, a  mixture of systems may be used for query processing as we show later in Section~\ref{sec-results}.

The importance of each metric may be weighed differently for a particular environment.
%Different weights may be preferred for each metric in a given environment.
The purpose of including all five of them is to help understand the impact
of various tradeoffs in order to guide system design.
Next we describe the specifics of our workload and how 
to evaluate system performance using these metrics.

%% file: sec-three.tex
\section{The Workload}\label{sec-three}

Our workload considers 8 hypothetical analysts who write queries
for marketing scenarios involving restaurants using social media data and static data.
For social media data we use a sample of the Twitter data stream and user check-in data from Foursquare.  
For static data we include a Landmarks data set (landmark locations).
Each analyst poses one query which is then revised multiple times.
%representing iterative exploratory process.
There are 4 versions of each query, representing the original query and 
3 subsequent revisions.
Next we define the types of changes allowed for each revision, and then provide 
a workload that uses these changes.

\subsection{Query building blocks}\label{sec-query-building-blocks}

Queries that evolve during exploratory data analysis may follow certain patterns of common changes.
As an analyst revises a query, she may tweak the selectivity to produce greater or fewer answers, include additional data sources for
stronger evidence of hypothesis, add a UDF to perform a specialized processing function, or refine the results by including or
removing a query sub-goal as more is learned about the data after each query revision.

To make these changes concrete, we evaluated complex analytical queries from several sources
to find evidence of the manner in which queries evolve.
The TPC-DS~\cite{tpc} workload includes 4 interactive OLAP queries that go through 2 revisions each.
Taverna~\cite{taverna} queries on MyExperiment~\cite{myexperiment} are scientific queries that retain all of their revisions.
Each of the top 10 most-downloaded Taverna queries had 2--11 revisions.
Yahoo!~Pipes~\cite{yahoo} has many versions of user queries over open-access web data, with more than 99 data sources.
Queries in Pipes are easily clone-able and modified by any user, and 
in one instance we observed a query with more than 49 revisions.
These observations suggest that queries in evolutionary analytics typically go through several revisions. % before becoming stable.  
%At that point, the query may become part of the standard business intelligence (BI) and is then included in the reporting workload.

Specifically, we commonly observed the following 4 types of changes during query revisions from a sampling of~\cite{taverna,tpc,yahoo}.
We note these changes are not mutually exclusive nor exhaustive but representative.

%Below we identify the common changes that an analyst may perform from one
%revision to another, and use these as building blocks for each evolving query.

\onefig{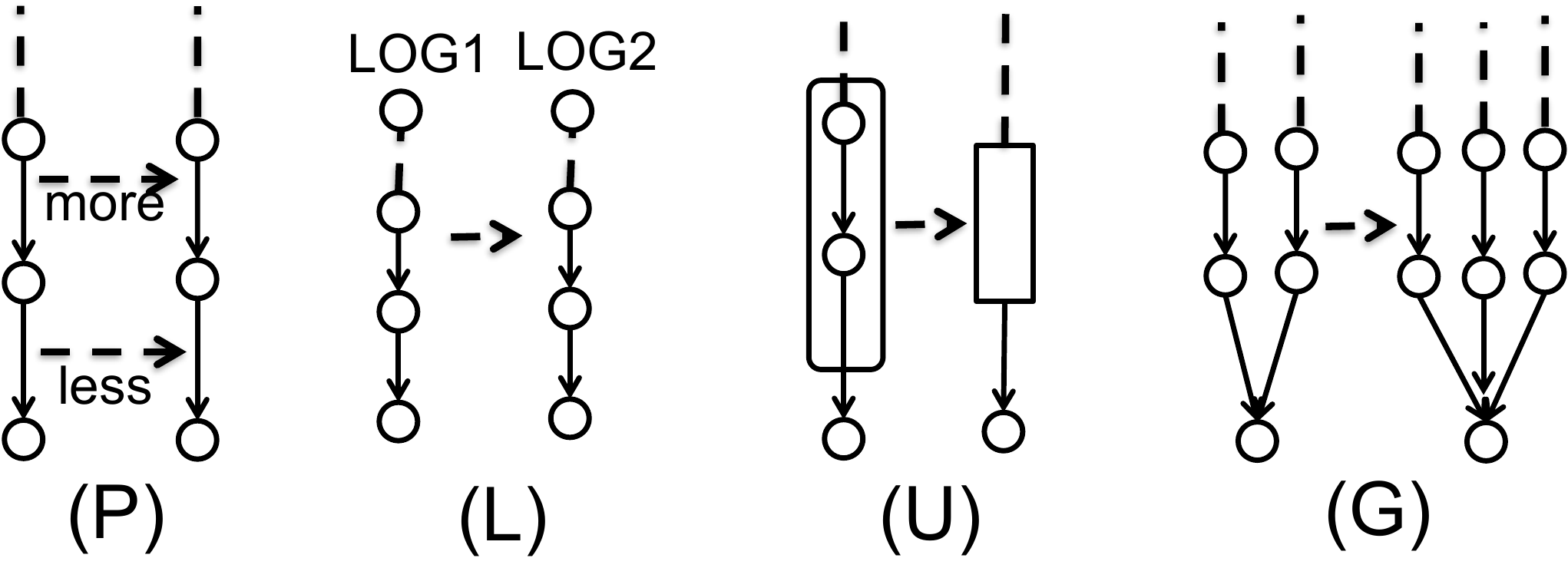}
{Dimensions of change for query revisions}{2.2in}

(P) Parameters: The query parameters are modified to obtain slightly
 different results (Figure~\ref{fig-slug-illustration}P).
For example, the analyst may alter a selection predicate or a top-$k$ value to
allow more or less data in the output.

(L) Logs: An analyst may make use of an additional data source in the query
 to obtain richer results (Figure~\ref{fig-slug-illustration}L).

(U) UDFs: An analyst may add or replace a set of operations in the query
 with a specialized UDF.

(G) Sub-Goals: Typically, an analyst writes several sub-queries
that each achieves a single goal and then joins these to obtain
the final output (Figure~\ref{fig-slug-illustration}G).
A revision may add or remove a sub-goal.
%For instance, in Example~\ref{ex-query-a2v2}  the analyst may add an additional
%sub-goal that specifies the user is also tech-savvy.

These four dimensions \{P,L,U,G\} serve as our evolving query building blocks.
For each query revision, the changes 
%from one version to the next 
are expressed by one or more of these dimensions.

\input{scenario-plug}

\subsection{Queries}
We give a high-level description of each query scenario in 
Table~\ref{tab-experiment-scenarios} left column, and the right column
 specifies the change from one version to the next in terms of our query building blocks.
For instance, let $Q$ represent the analyst's first version of the query.
Then each subsequent version (i.e., 2,3,4) is represented by indicating the dimensions
that were changed during each revision in the following way:

\vspace{-0.1in}
\begin{equation}
Q \rightarrow \{P,L,G\} \rightarrow \{P\} \rightarrow \{P,G\}
\end{equation}

To illustrate this process, we start with Example~\ref{ex-query-v1} 
as the first version of Analyst 1's query in Table~\ref{tab-experiment-scenarios}.
This version is indicated by $Q$ above.
%The analyst's task for query 1 is: \textit{Identify likely customers via 
%social media data for a marketing  campaign of a new wine being introduced 
%in the Bay Area}.
Analyst 1 first desires to find users who like wine,  are affluent,
and have many good friends.

\begin{example}\label{ex-query-v1}

(a): EXTRACT user from Twitter log.
Apply UDF-CLASSIFY-WINE-SCORE on each user tweet to obtain a wine-score.
Groupby user, compute a wine-sentiment-score for each user.

(b): From Twitter log, apply UDAF-CLASSIFY-AFFLUENT on tweets to classify a user 
as affluent or not.

(c): From Twitter log, create social network between every user pair 
using tweet source and dest.
GROUPBY user pair in social network, count tweets. 
Assign friendship-strength-score to each user pair.

JOIN (a),(b), and (c).
Threshold based on wine-sentiment-score, friendship-strength-score.
\end{example}

Next the analyst wants to find more evidence that the user likes wine.
She revises the query by changing \{P,L,G\}, adding two new data sources 
 \{$L$\} (Foursquare and Landmarks), 
a new sub-goal \{$G$\} that computes a checkin-count for users who
go to wine places, and decreases the threshold parameter \{$P$\} for wine-sentiment-score
since she will have evidence a user likes wine from 2 data sources.
Example~\ref{ex-query-a2v1} below describes version 2 of the query.

\begin{example}\label{ex-query-a2v1}  % *********** A2v1

(d): EXTRACT from Foursquare log.
For each checkin, obtain the user and restaurant name.
Using the Landmarks data, filter by checkin to places of type wine-bar.
Groupby user, compute checkin-count.

(e): Decrease wine-sentiment-score threshold

JOIN (a),(b),(c) and (d).
Threshold based on new wine-sentiment-score in (e),
friendship-strength-score.
\end{example}

Query versions 3 and 4 are revised similarly but are omitted 
due to lack of space.
A detailed description of all queries is provided in the extended version~\cite{lefe2013b}.
%The extended version of this paper~\cite{lefe2013b} provides an appendix 
%describing all queries.

%% file: scenario-plug.tex
\begin{table*}[!t]
\caption{\label{tab-experiment-scenarios}Eight analyst marketing scenarios,
along with the dimensions modified during each of the 4 evolutions}
\small{
\centering
\begin{tabular}{| p{0.70\linewidth} | p{0.28\linewidth}|}
\hline

Analyst1 wants to identify a number of ``wine lovers'' to
send them a coupon for a new wine being introduced in a local region.
This evolution investigates ways of finding suitable users to whom
sending a coupon would have the most impact.
&
$Q \rightarrow \{P,L,G\} \rightarrow \{P\} \rightarrow \{P,G\}$

\\\hline

Analyst 2 wants to find influential users who 
visit a lot of restaurants for inclusion in an advertisement campaign.
The evolution of this scenario will focus on increasingly sophisticated ways
of identifying users  who are ``foodies''.
%Analyst 2 wants to find influential users who are
%interested in food for an advertisement campaign.
%The evolution of this scenario includes increasingly
%sophisticated interpretations of what it means to be ``influential''.
&
$Q \rightarrow \{L,U,G\} \rightarrow \{P,G\} \rightarrow \{P,G\}$
\\ \hline

Analyst3 wants to start a gift recommendation service where friends
can send a gift certificate to a user $u_1$.
We want to generate a few restaurant choices based on $u_1$'s preferences and his friend's preferences.
The evolution in this scenario will investigate how to
generate a diverse set of recommendations that would cater to $u$ and
his close set of friends.
%Analyst3 wants to start a gift recommendation service where friends
%can send a gift certificate to a user $u$.
%We want to generate a few restaurant choices based on $u$'s preferences.
%The evolution in this scenario will investigate how to
%generate a diverse set of recommendations that would cater to $u$ and
%his close set of friends.
&
$Q \rightarrow \{P,G\} \rightarrow \{P,L,G\} \rightarrow \{G\}$
\\ \hline

Analyst4 wants to identify a good area to locate a sports bar.
The area must have a lot of people who like sports and check-in to bars,
but the area does not already have too many sports bars
in relation to other areas.
The evolution focuses on identifying a suitable area where
there is high interest but a low density of sports
bars.
% Analyst4 wants to provide restaurants with a breakdown of their
%customer population.
%In particular, the analyst wants to provide breakdowns in terms of who goes to a
%particular restaurant.
%The analyst may also want to perform analysis in terms for geographical area.
%For instance, for any area which are the popular restaurants by categories etc.
%The goal of this scenario is to perform increasingly complicated analytical
%operations on the log.
&
$Q \rightarrow \{U,G\} \rightarrow \{L,U,G\} \rightarrow \{U,G\}$
\\ \hline

Analyst5 wants to give restaurant owners a customer poaching tool.
For each restaurant $r$, we identify customers who go to a ``similar''
restaurant in the area but do not visit $r$.
The owner of $r$ may use this to target advertisements.
The evolutionary nature focuses on determining ``similar'' restaurants and their users.
&
$Q \rightarrow \{L,G\} \rightarrow \{L,U\} \rightarrow \{P,G\}$
\\ \hline

Analyst6 tries to find out if restaurants are losing loyal customers.
He wants to identify those customers
who used to visit more frequently but are now visiting other restaurants in the area
so that he can send them a coupon to win them back.
The evolutionary nature of this scenario will focus on
how to identify prior active users.
&
$Q \rightarrow \{L,G\} \rightarrow \{P,G\} \rightarrow \{P,G\}$
\\ \hline

Analyst7 wants to identify the direct competition for poorly-performing restaurants.
He first tries to determine if there is a more successful restaurant
of similar type in the same area.
The evolutionary nature focuses on identifying good and bad restaurants in an area,
as well as what customers like about the menu, food, service, etc. about the successful
restaurants in the area.
&
$Q \rightarrow \{L,G\} \rightarrow \{G\} \rightarrow \{U,G\}$
\\ \hline

%Analyst8 wants to identity what category (i.e., american, indian,
%chinese, steak) of restaurants is missing from a
%certain region. For example, one of the outcome can be that
%Cupertino could really use a Moroccan restaurant.
%The evolutionary nature of this scenario will focus on how to
%find out if there is a need for a good restaurant in a certain
%category and if so what is the nature of their clientele etc.
%&
%$Q \rightarrow \{x,x,x\} \rightarrow \{x,x,x\} \rightarrow \{x,x,x\}$
%\\ \hline

Analyst8 wants to recommend a high-end hotel vacation in an area users will
like based on their known preferences for restaurants, theaters, and luxury items.
The evolutionary nature focuses on matching user's
preferences with the types of businesses in an area.
%Analyst8 wants to recommend a high-end hotel vacation to affluent people.
%This involves finding users who like luxury items
%and like to travel, and identifying the types of locations such as
%restaurants or theater these clientele frequent in their local areas.
%Then the analyst determines the density of these same types
%of locations per global cell grid.
%The analyst can then send a coupon to those users for a high-end
%hotel with a high density of nearby activities that user would like.
%The evolutionary nature of this scenario is how to identify
%users who like luxury items and events and find a grid cell location
%with a luxury hotel that has these same activities for a given user.
&
$Q \rightarrow \{L,G\} \rightarrow \{U,G\} \rightarrow \{P,L,U,G\}$
\\ \hline
\end{tabular}
}
\vspace{-0.2in}
\end{table*}

%% file: sec-four.tex
\section{Running the Benchmark}\label{sec-four}

%base system state is when no queries have been executed.
We now present our benchmark methodology for query evolution, user evolution, and data evolution and we  show an example of benchmark results.
We consider the initial system state to be \textit{idle}, with no previously loaded data or executed queries.
%For each test below, all metrics should be recorded and the specified comparisons performed.
%In each test, the workload is observed online, it is not provided offline.

\subsection{Benchmark methodology}

\paragraph*{Query evolution}
This test will use all analysts 1--8 and all query versions from each analyst.
%The goal of this test is to measure system performance for the case of an analyst executing all versions of his query. 
(1) From initial system state, execute analyst 1 query versions 1 through 4 in succession, returning to initial system state before each version.
(2) From initial system state, execute analyst 1 query versions 1 through 4 in succession, without returning to initial system state before each version.
%Return to initial system state before query version 1 only.  
Compare metrics from (1) and (2), and repeat for each remaining analyst.
This comparison highlights a system's ability to process any
repeating tasks from the same user.

\paragraph*{User evolution}
%The goal of this test is to measure how well the system supports the introduction of new users.
This test will use all analysts 1--8 but only version 1 of each analyst's query.
%For this test we will only use version 1 of each analyst's query.
First, assume some order of analysts 1--8.  
(1) From initial system state, execute each analyst's query in the chosen order, returning to initial system state before each query. 
%returning to initial system state before each query. %from the initial system state each time.
(2) From initial system state, execute each analyst's query in the chosen order, without returning to initial system state before each query.
%Starting from the initial system state, execute each analyst's query in the chosen order.
Compare metrics from (1) and (2).
This comparison highlights a system's ability to process 
similar tasks from different users.

\paragraph*{Data evolution}
%The goal of this test is to measure the system effort required to enable analysts to access new data. 
This test will use a single data source, e.g., Twitter log, and the subset of data requested in the first step should be 
a number of columns equal to half of the total number of columns in the log schema. 
The columns should be randomly chosen each time.
(1) From an initial system state, an analyst requests a subset of data from a new data source.  
(2) An analyst requests one \textit{additional} attribute from the data source in (1), in each successive version of the query.
(3) A new analyst requests a subset of data previously accessed by the analyst in (1).
(4) Repeat (1), (2), (3) returning to the initial state after each query.
Compare metrics from  (1), (2), (3) with (4).
This comparison highlights a system's ability to access subsets of data from a new data source on demand.

\onefigfullrow{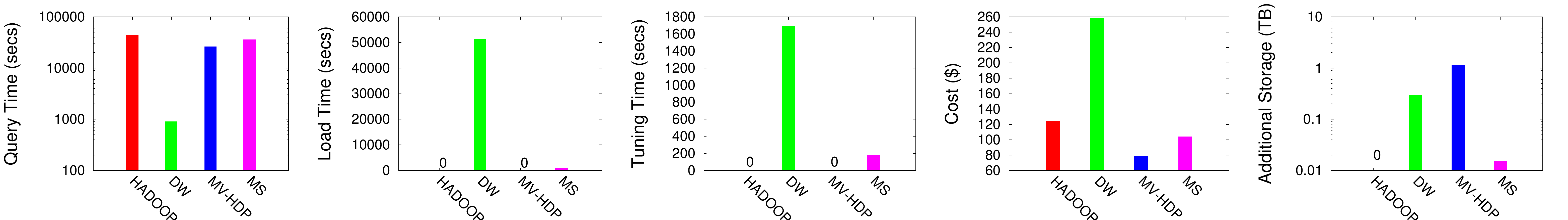}{Figure shows a sample reporting of 4 data systems on a user evolution scenario}{7in}

\subsection{Example benchmark results}\label{sec-results}

Next, we briefly show a sample reporting on
the relative performance of four data systems using our workload and metrics
for a user evolution scenario.
%We evaluate the performance of the four systems by executing the first versions of each of the 8 analysts queries one at a time.
The experimental setup consists of 9 nodes running a widely used commercial parallel data warehouse (DW) and 14 nodes running Hadoop.
The ratio of Hadoop nodes to DW nodes is 1.5$\times$. 
The DW and Hadoop clusters are independent, and nodes are connected with 1 GbE.
Each node has two 2.4 GHz \textsc{xeon} CPUs and a local 2 TB disk.
In this test, our data includes a 1 TB Foursquare log, a 1 TB  Twitter log and 12 GB Landmarks log.
We use Hive~\cite{thus2010} to execute our queries on the Hadoop system.
Since all systems utilize the base data stored in Hadoop,  we omit this from the storage metric.
%For every experimental scenario, the base data is initially present in Hadoop.
%Since this is common in each system tested, we omit this from the storage metric.

Figure~\ref{fig-ue-plot} reports the results for the user evolution scenario.
{\sf HADOOP} corresponds to a Hadoop-only execution of the query.
{\sf DW} executes the query on the DW but uses Hadoop as an ETL tool
to extract the subset of data required by the query.
{\sf MV-HDP} corresponds to a system that we developed in~\cite{lefe2013}
that rewrites Hadoop queries based on opportunistic materialistic views 
left behind from prior execution runs.
{\sf MS} is an implementation of a multi-store query optimizer
(similar to~\cite{simi2012}) that uses both Hadoop and a data warehouse 
together to execute each query.
For {\sf MS}, the load time refers only to the time to move and then load 
data on-the-fly from {\sf HADOOP} to {\sf DW} after partial execution 
in {\sf HADOOP}.

It can be seen from the figure that reporting on the 5 metrics 
exposes their tradeoffs, which are not easily
captured when reporting just on the query execution time.
Since Hadoop performs ETL on the fly, query performance is quite
poor compared to the DW.
On the other hand, the superior performance of the DW is 
offset by the  high cost of loading the data into the data warehouse.
Both {\sf MV-HDP} and {\sf MS} show tradeoffs that reduce query response time.
% better performance than {\sf HADOOP} and {\sf DW}, respectively.
{\sf HADOOP} and {\sf MV-HDP} do not incur any tuning overhead 
whereas {\sf DW} and {\sf MS} require a tuning phase to provide good
performance. 
We used Amazon EC2 and Redshift~\cite{amazon-aws}
 pricing to approximate the dollar cost of the machines and storage 
 (using cost for machines similar to those in our in-house clusters).
The cost values show that {\sf HADOOP} is far cheaper than DW while 
{\sf MS} is cheaper than both,
and {\sf MV-HDP} has the lowest cost.
%Finally, the storage corresponds to how much storage overhead each of the  system incurs over the base logs.
Finally, it can be seen that {\sf MV-HDP} incurs a significant storage overhead by retaining 
results as opportunistic views from all the prior executions runs.
The tradeoff with storage size improves query response time compared to {\sf HADOOP}.

%% file: sec-five.tex
\section{Discussion}\label{sec-five}

In the new analytical space, the key question is how to design systems
to address emerging needs.
The continued  popularity of Hadoop and data warehouses notwithstanding,
these are only suitable when the required use-case matches either of their starkly
different  characteristics.  
One focuses on being able to query the data right away, tolerating lesser performance.
The other focuses on performance at the expense of significant delay in being able to query the data. 
These systems represent two ends of a spectrum, and
the influx of so many new data processing systems shows that these two distinct choices are not meeting all current needs.
%However, clearly the influx of so many new data processing systems shows that these
%two distinct choices are not meeting all current needs.

In this paper, our metrics highlight the tradeoffs among many design choices
and the metrics can be used to guide system development.
For example, we show 2 systems that remedy one dimension by shifting 
the tradeoff with another dimension.
With {\sf MV-HDP} we show that increased storage leads to better performance than Hadoop.  
With {\sf MS} we show that one can remedy the loading time of a data warehouse to an extent
by sacrificing some of DW performance.
An interesting further research direction is to leverage
the best properties 
of several systems to create hybrid systems.
%If we learn to manage these tradeoffs better, it may be possible to design systems
%that adapt dynamically to changing user needs. 

%% file: appendix.tex
\normalsize
\section{Query Descriptions}

In this section we describe each of the queries.
% each Analyst's query and describe its versions.
These queries use three datasets:  a Twitter data stream of user tweets, 
 a Foursquare (4SQ) data stream of user checkins, and a Landmarks log of 
 locations and their types.
The identity of users is common across the Twitter and 4SQ logs, 
while  identity of locations is common across the 4SQ and Landmarks logs.
All logs are stored as \textsc{json} text.

We present 8 Analyst's queries with 4 versions each.
For each analyst, we state a high-level goal of what the analyst is trying to
achieve, 
as well as 4 query versions toward the stated goal.
Each query version modifies the previous version.
For each query, we describe the task rather than provide an implementation in particular language 
since many are possible, e.g., SQL, HiveQL, Pig, Java, etc.

During a typical exploration, and analyst may spend a lot of time identifying the  data distribution to get an understanding of where the density and sparsity lies.
For this reason, the queries below have parameters indicated by an 
\underline{underline} that an analyst would modify in order to obtain a 
representative answer set.
Choosing an appropriate value or interpretation of the underlined parts of
the queries is a necessary step an analyst performs.
This may require a trial and
error process  resulting in additional versions of the queries.
We leave these values unspecified as they are a function of the real-world datasets used.

Finally, it is important to note that the queries are not rigid interpretations of a goal, but rather
one approach toward answering a question.
For example, there may be several ways to interpret what it means to be
``good'' friends in a social network.
Furthermore, even for a particular interpretation, there may be several ways to 
express it as a query. 
Hence, other variations of a query are possible.

\subsection*{User Defined Functions referenced by queries}

The following queries reference multiple UDFs, for which we provide a brief description below.
Text classifiers can be implemented using a \textit{bag-of-words} method.
For example, classes such as \textsc{coffee-drinker} may include the words 
(\textit{coffee, espresso, latte, french press}) while \textsc{wine-lover} 
may include (\textit{cabernet, vineyard, merlot, chardonnay}).
By accepting a bag of words as an argument, these UDFs are easily reusable.
However, this does not exclude other classification methods.
UDAFs perform a groupby on a key and apply similar classification on the elements of a group.

\begin{enumerate}
\item UDF-CLASSIFY-WINE-SCORE: Input is text and output is wine-score indicating a strong presence of wine-terms.

\item UDF-CLASSIFY-FOOD-SCORE: Input is text and output is food-sentiment-score indicating a strong presence of food-terms.

\item UDF-GRID-CELL:  Input is lat-lon coordinates and grid resolution, and output is a grid-cell number.

\item UDF-CLASSIFY-BEER-SCORE: Input is text and output is beer-score indicating a strong presence of beer-terms.

\item UDF-MENU-SIMILARITY:  Input is two lists of menu items and output is a score indicating the similarity of the lists.

\item UDF-NLP-ENTITITY-SENTIMENT:  Input is text and output is the entities extracted from the text with a sentiment-score for each entity.

\item UDF-CLASSIFY-LUXURY-SCORE:  Input is text and output is a binary value indicating if the text concerns luxury items. 

\item UDF-SENTIMENT:  Input is text and output is a sentiment-score expressing positive or negative sentiment with the score indicating the strength of the sentiment.

\item UDAF-CLASSIFY-AFFLUENT:  Input is all text from a given user and output is a binary value indicating if the user is affluent or not.

\item UDAF-CLASSIFY-SPORTS: Input is all text from a given user and output is a binary value indicating if the user is interested in sports or not.
\end{enumerate}

\subsection{Analyst 1}

Analyst1 wants to identify a number of ``wine lovers'' to
send them a coupon for a new wine being introduced in a local region.
This evolution investigates ways of finding suitable users to whom
sending a coupon would have the most impact.

\subsubsection{Analyst 1, Version 1}

\begin{itemize}
  \item Analyst goal: Find users that like wine, have strong friendships, and are affluent.

  \item Query:   From Twitter, apply UDF-CLASSIFY-WINE-SCORE on each user's tweets and groupby user to produce wine-sentiment-score for each user.
  Threshold on wine-sentiment-score above \underline{$x_1$}.   
  
  From Twitter, compute all pairs $\langle u_1,u_2\rangle$ of users that communicate
  with each other, assigning each pair a friendship-strength-score based on the number of times they communicate.
  %Select user pairs having friendship-strength-score greater than \underline{value1}.
  Threshold on friendship-strength-score above \underline{$x_2$}.
  
  From Twitter, apply UDAF-CLASSIFY-AFFLUENT on users and their tweets.

  Join results by user.
\end{itemize}

\subsubsection{Analyst 1, Version 2}

\begin{itemize}
  \item Analyst goal: Next, consider users to be wine-lovers if they checkin to many wine places. 

  \item Query:   From previous version, reduce wine-sentiment-score threshold to \underline{$x_2'$} 
  since now there will be additional evidence a user likes wine.
  
  From 4SQ, identify places that users checkin.
  Join with places in Landmarks.
  Select users that checkin to places of type wine-bar.  
  For each user, count the number of checkins.
  Threshold on checkin-count above \underline{$x_3$}.
  
  Join these with users from previous version.
  
\end{itemize}

\subsubsection{Analyst 1, Version 3}

\begin{itemize}
  \item  Analyst goal:  Now find users that are also in the San Francisco area as well 
  as prolific on Twitter.

  \item Query: 
  From previous version, select users \underline{local to} San Francisco.
  Threshold on tweet-count above \underline{$x_4$}.
  Adjust  $x_1$, $x_2$, $x_3$ appropriately to produce ``enough'' answers. 
  % e.g., more than 20% of tweets about the location mention a wine-term.
\end{itemize}

\subsubsection{Analyst 1, Version 4}

\begin{itemize}
  \item Analyst goal: Finally, require that a user's friends must also visit wine-places.

  \item  Query: For user pairs $\langle u_1,u_2\rangle$, 
  threshold on $u_2$ checkin-score above \underline{$x_5$}.
  For each user $u_1$, count the number of friends with checkin-count above threshold.
  Retain $u_1$ if count above \underline{$x_6$}.  
  Join these with users from previous version.
 
\end{itemize}

\subsection{Analyst 2}

Analyst 2 wants to find influential users who 
visit a lot of restaurants for inclusion in an advertisement campaign.
The evolution of this scenario will focus on increasingly sophisticated ways
of identifying users  who are ``foodies''.

\subsubsection{Analyst 2, Version 1}

\begin{itemize}
\item Analyst goal: Find users who frequently visit restaurants.

\item Query:   From 4SQ, identify places that users checkin.
  Join with places in Landmark log.
  Select users that checkin to places of type restaurant. 
  For each user, count the number of times they checkin to a place of type restaurant.
  Compute the normalized-count based on the maximum count across all users.
  Threshold on normalized-count above \underline{$x_1$}.
\end{itemize}

\subsubsection{Analyst 2, Version 2}

\begin{itemize}
\item Analyst goal: Additionally, user also likes food if they talk positively about food. 

\item Query:
  From Twitter, apply UDF-CLASSIFY-FOOD-SCORE on each user's tweets and groupby user to produce food-sentiment-score for each user.
  Threshold on food-sentiment-score above \underline{$x_2$}. 
  
  Join these users with users in previous version.
\end{itemize}

\subsubsection{Analyst 2, Version 3}

\begin{itemize}
\item Analyst goal: Further define that a user likes food if they dine at many different types of restaurants.

\item Query: Revise previous version by counting the number of times a user has visited each distinct type of restaurant.
Select users who have visited \underline{$x_3$} distinct restaurant types at least \underline{$x_4$} times.
\end{itemize}

\subsubsection{Analyst 2, Version 4}

\begin{itemize}

\item Analyst goal:  Finally, require that these users do not frequently visit
restaurants with low ratings.  

\item Query:  From previous version, compute the percentage of each user's checkins
to restaurants with ratings less than \underline{$x_5$}.
Threshold on percent below \underline{$x_6$}. 
\end{itemize}

\subsection{Analyst 3}
Analyst3 wants to start a gift recommendation service where friends
can send a gift certificate to a user $u$.
We want to generate a few restaurant choices based on $u$'s preferences and $u$'s friend's preferences.
The evolution in this scenario will investigate how to
generate a diverse set of recommendations that would cater to $u$, and $u$'s close set of friends.

\subsubsection{Analyst 3, Version 1}

\begin{itemize}
  \item Analyst goal: For each user $u$, identify those restaurants that $u$'s good friends frequently visit.
  
  \item Query: From Twitter, compute all pairs $\langle u_1,u_2\rangle$ of users that communicate
  with each other, assigning each pair a friendship-strength-score based on the number of times they communicate.
  Threshold on friendship-strength-score above \underline{$x_1$}.
  
  From 4SQ, identify places that users checkin.
  Join with places in Landmarks.
  Select users that checkin to places of type restaurant. 
  
  For each user $u_1$, find all the restaurants that her friends $u_2$ have visited.
  For each restaurant, count the number of checkins.
  Threshold on count above \underline{$x_2$}. 
\end{itemize}

\subsubsection{Analyst 3, Version 2}

\begin{itemize}
  \item Analyst goal:   Next, only consider users that have friends in the same area as well as 
  other friends in common.
  
  \item Query: Revise the previous version by redefining what it means to be
  good friends.  From Twitter, recompute all pairs $\langle u_1,u_2\rangle$ of users that live
  in the \underline{same} area, and have a friendship-strength-score above \underline{$x_3$}.
  Additionally, a user pair $\langle u_1,u_2\rangle$ are said to be good friends if
  they have more than \underline{$x_4$} friends in common.
  
\end{itemize}

\subsubsection{Analyst 3, Version 3}

\begin{itemize}
  \item Analyst goal: Next, identify only those restaurants that are the same type as
  a user's favorite restaurant.

  \item Query: From 4SQ, for each user $u_1$, find favorite restaurant type 
  by counting the number of checkins to each restaurant, and select the restaurant with the
  max number of checkins as $u_1$'s favorite restaurant $r$.
  
  Join with Landmarks to obtain $r$'s  type. 
  
  From the previous version, select only those restaurants for $u_1$ that belong to the same type as $u_1$'s favorite type.
  
\end{itemize}

\subsubsection{Analyst 3, Version 4}

\begin{itemize}
  \item Analyst goal:  Finally, find additional restaurants that are similar 
  to those visited by a user's friends. 
  
  \item Query: From 4SQ, for all restaurant pairs $\langle r_1,r_2\rangle$, 
  count the number of users that have visited both restaurants.
  Threshold on count above \underline{$x_5$}.
  All remaining pairs $\langle r_1,r_2\rangle$ are considered to be similar  since they have many common customers. 
  For each user $u_1$, suggest $r_2$ if $r_1$ is a restaurant frequently visited
  by $u_1$'s friends.
  
\end{itemize}

\subsection{Analyst 4}

Analyst4 wants to identify a good area to locate a sports bar.
The area must have a lot of people who like sports and check-in to bars,
but the area does not already have too many sports bars
in relation to other areas.
The evolution focuses on identifying a suitable area where
there is high interest but a low density of sports
bars.

\subsubsection{Analyst 4, Version 1}

\begin{itemize}
\item Analyst goal: Find users who like beer and where they live.

\item Query: From 4SQ, identify users and their location that frequently mention the word
``beer'' in their text.  
For each user, count the occurrences of the word. 
Threshold on count above \underline{$x_1$}.

\end{itemize}

\subsubsection{Analyst 4, Version 2}

\begin{itemize}
\item Analyst goal: Next, find areas where there are many beer lovers.

\item Query: From the previous version, use UDF-GRID-CELL to map user locations to a grid cell.
Count number of users in each grid cell.
Threshold on count above \underline{$x_2$}.

\end{itemize}

\subsubsection{Analyst 4, Version 3}

\begin{itemize}
\item Analyst goal: Next, find areas with many users that like beer and sports
but do not have many sports bars.  

\item Query:
 From Twitter, apply UDAF-CLASSIFY-SPORTS on users and their tweets.
 Then apply a UDF-CLASSIFY-BEER-SCORE to better identify users that like beer,
 and produce a beer-score for each user.
 Join sports and beer users.
 Threshold on beer-score above \underline{$x_3$}.
 Next, apply UDF-GRID-CELL to map user locations to a grid cell.
 Count number of users in each grid cell.
 Threshold on count above \underline{$x_4$}.

From Landmarks, obtain restaurant name, type and location.
Select places that are type equal to sports bar.
 Next, apply UDF-GRID-CELL to map place locations to a grid cell.
 Count number of restaurants in each grid cell.
 Threshold on count below \underline{$x_5$}.
 
 Join grid cells from user locations and sports bar locations.
\end{itemize}

\subsubsection{Analyst 4, Version 4}

\begin{itemize}
\item Analyst goal: Finally, find area with high user interest but 
few popular sports bars relative to the number of users.

\item Query:  From 4SQ, identify places that users checkin.

  Join with places in Landmarks.
  Select places that are type equal to sports bar.
  For each place, count the number of checkins.
  Threshold on count above \underline{$x_6$}.
  
  Next, apply UDF-GRID-CELL to map place locations to a grid cell.
  Count the number of places per grid cell.

  Join this with the grid cells from previous version.
  Threshold on ratio of user to sports bars count above \underline{$x_7$}.
  
\end{itemize}

\subsection{Analyst 5}

Analyst5 wants to give restaurant owners a customer poaching tool.
For each restaurant $r$, we identify customers who go to a ``similar''
restaurant in the area but do not visit $r$.
The owner of $r$ may use this to target advertisements.
The evolutionary nature focuses on determining ``similar'' restaurants and their users.

\subsubsection{Analyst 5, Version 1}

\begin{itemize}
  \item Analyst goal: Find similar restaurants based the overlap of users that checkin
  to each place.
  
  \item Query: From 4SQ, for all restaurant pairs $\langle r_1,r_2\rangle$, 
  count the number of users that have visited both restaurants.
  Threshold on count above \underline{$x_1$}.
  
\end{itemize}

\subsubsection{Analyst 5, Version 2}

\begin{itemize}
  \item Analyst goal: Next, find restaurants that are similar as indicated by a user or the user's friends
  frequently visiting the same places.

  \item Query: From Twitter, compute all pairs $\langle u_1,u_2\rangle$ of users that communicate
  with each other, assigning each pair a friendship-strength-score based on the number of times they communicate.
  Threshold on friendship-strength-score above \underline{$x_2$}.
  
  From 4SQ, for all restaurant pairs $\langle r_1,r_2\rangle$, 
  count the number of users that have visited both restaurants, as well as the number of times a user $u_1$ 
  has visited $r_1$ and one of their friends $u_2$ has visited $r_2$.
  Threshold on count above \underline{$x_3$}.
\end{itemize}

\subsubsection{Analyst 5, Version 3}

\begin{itemize}
  \item Analyst goal:  Next, find restaurant pairs that are also similar based on the
  similarity of their menus.
  
  \item Query: From Landmarks, create restaurant pairs $\langle r_1',r_2'\rangle$  
  that have the same zip code and type.
  For each pair, apply UDF-MENU-SIMILARITY to obtain menu-similarity-score.  
  Threshold on menu-similarity-score above \underline{$x_4$}.
  
  Join pairs $\langle r_1',r_2'\rangle$ with pairs $\langle r_1,r_2\rangle$ from the previous version.

\end{itemize}

\subsubsection{Analyst 5, Version 4}

\begin{itemize}
  \item Analyst goal:  Finally, find users that visit one restaurant but not a similar restaurant.
  
  \item Query: From 4SQ, for each restaurant $r$, identify the users that have visited $r$ 
  and the count of times they have visited.
  For each restaurant pair $\langle r_1,r_2\rangle$ from the previous version, 
  select users $u$ that have visited $r_1$ more than \underline{$x_5$} times and visited $r_2$ less than \underline{$x_6$} times.

\end{itemize}

\subsection{Analyst 6}
Analyst6 tries to find out if restaurants are losing loyal customers.
He wants to identify those customers
who used to visit more frequently but are now visiting other restaurants in the area
so that he can send them a coupon to win them back.
The evolutionary nature of this scenario will focus on
how to identify prior active users.

\subsubsection{Analyst 6, Version 1}

\begin{itemize}
\item Analyst goal: For each restaurant, identify other restaurants with the same zip code and type that are less popular. 

\item Query: From Landmarks, create restaurant pairs $\langle r_1,r_2\rangle$  
  that have the same zip code and type, and $r_2$ has a \underline{much} lower checkin count than $r_1$.

\end{itemize}

\subsubsection{Analyst 6, Version 2}

\begin{itemize}
\item Analyst goal: Now, identify restaurants that have lately become
less popular.

\item Query:   From 4SQ, identify places that users checkin.

  Join with places in Landmarks that have type restaurant.
  For each restaurant,
  compute the average number of checkins per month in the last \underline{$x_1$} months
  and the number of checkins in the last 1 month.  
  
  Threshold on the ratio of recent checkins to historical average checkins below \underline{$x_2$}.
  
\end{itemize}

\subsubsection{Analyst 6, Version 3}

\begin{itemize}
\item Analyst goal:  Next, find users that stopped visiting those restaurants.

\item Query: For  restaurant $r$ identified as becoming less popular in the previous version,
and a user $u$ that visited $r$,
compute the average number of checkins per month by user $u$ in the last \underline{$x_3$} months
  and the number of checkins by user $u$ in the last 1 month.
  Compute the ratio of recent checkins to historical average checkins
  
    Threshold on  ratio below \underline{$x_4$}.
\end{itemize}

\subsubsection{Analyst 6, Version 4}

\begin{itemize}
\item Analyst goal: Finally, find users that no longer frequent a particular restaurant 
but still visit other  restaurants in the same area.

\item Query:  From 4SQ, for each user $u$, count the number of checkins by zip code.
   Threshold on count above \underline{$x_5$}.
   
   For each less popular restaurant $r$ identified in previous version,
   retain $u$ only if $u$ still frequently visits restaurants in the same zip code as $r$.
\end{itemize}

\subsection{Analyst 7}
Analyst7 wants to identify the direct competition for poorly-performing restaurants.
He first tries to determine if there is a more successful restaurant
of similar type in the same area.
The evolutionary nature focuses on identifying good and bad restaurants in an area,
as well as what customers like about the menu, food, service, etc. about the successful
restaurants in the area.

\subsubsection{Analyst 7, Version 1}

\begin{itemize}
\item Analyst goal: For each zip code, identify good and bad restaurants.

\item Query: From Landmarks, identify places that are restaurants, 
and apply UDF-SENTIMENT on the restaurant comments to obtain a sentiment-score.
For each zip code, retain restaurants with sentiment-score above \underline{$x_1$} or below \underline{$x_2$}
as good and bad restaurants.

\end{itemize}

\subsubsection{Analyst 7, Version 2}

\begin{itemize}
\item Analyst goal:  Next,  refine the discrimination of restaurants as good and bad based on their popularity.

\item Query: From 4SQ, obtain the checkin count for every restaurant.

Threshold on count above \underline{$x_3$}.
Join with the good restaurants from the previous version.

Threshold on count below \underline{$x_4$}.
Join with the bad restaurants from the previous version.

\end{itemize}

\subsubsection{Analyst 7, Version 3}

\begin{itemize}
\item Analyst goal: Further discriminate restaurants as good and bad based repeat checkins.

\item Query: From 4SQ, obtain the checkin count for every restaurant.
For each restaurant, count the number of users that checkedin only once,
and the number of users that checkedin more than \underline{$x_5$} times.
Compute the ratio of single checkins to multiple checkins.

Threshold on ratio below \underline{$x_6$}.
Join with the good restaurants from the previous version.

Threshold on ratio above \underline{$x_7$}.
Join with the bad restaurants from the previous version.

\end{itemize}

\subsubsection{Analyst 7, Version 4}

\begin{itemize}
\item Analyst goal: Next, for each restaurant, find the most frequent entities 
with positive and negative comments.

\item Query: From 4SQ, apply UDF-NLP-ENTITITY-SENTIMENT per user checkin.
For each restaurant and each entity, aggregate the sentiment-score.
Threshold on sentiment-score above \underline{$x_5$}.

Join with good and bad restaurants from previous version.

\end{itemize}

\subsection{Analyst 8}

Analyst8 wants to recommend a high-end hotel vacation in an area users will
like based on their known preferences for restaurants, theaters, and luxury items.
The evolutionary nature focuses on matching user's
preferences with the types of businesses in an area.

\subsubsection{Analyst 8, Version 1}

\begin{itemize}
  \item Analyst goal:  Find users who talk about `luxury items'.

  \item Query: From Twitter, apply UDF-CLASSIFY-LUXURY-SCORE on user tweets.
  For each user, count the number of tweets about luxury-items.
  Threshold on count above \underline{$x_1$}.
  
\end{itemize}

\subsubsection{Analyst 8, Version 2}

\begin{itemize}
  \item Analyst goal:  Next, identify restaurants those users frequently visit.

  \item Query: From 4SQ, for each user, count the number of checkins per restaurant. 
  Threshold on count above \underline{$x_2$}. 
  
  Join with users from previous version.
  
  For each restaurant, count the total number of checkins by all these users.
  Threshold on count above \underline{$x_3$}.
  
\end{itemize}

\subsubsection{Analyst 8, Version 3}

\begin{itemize}
  \item Analyst goal: Next, find areas that have a high density of these restaurants and identify
  the distribution of restaurant types in the area.
  
  \item Query:  For the restaurants from the previous version, apply UDF-GRID-CELL.
  Count the number of restaurants per grid cell.
  Threshold on count above \underline{$x_4$}.
  
  For each grid cell, compute a histogram on the restaurant type and count.

\end{itemize}

\subsubsection{Analyst 8, Version 4}
\begin{itemize}
  \item Analyst goal:  Finally, match users to grid cells and find luxury hotels in their matching grid cell.

  \item Query: For each user $u$ from previous version, identify location, and 
  compute a histogram on the restaurant type and $u$'s checkin count.
  
  Match $u$ to a grid cell such that the grid cell is \underline{sufficiently} far away from $u$'s 
  location, and there is a \underline{significant} overlap between $u$'s histogram and grid-cell $g$
  histogram from previous version.

  From Landmarks, find hotels with rating greater than \underline{$x_5$} stars, 
  and apply UDF-GRID-CELL to convert hotel location to grid cell $g'$.
  
  For each user $u$, join grid cell $g$ with $g'$ to identify hotels in an area matching $u$'s restaurant preferences.

\end{itemize}